\documentclass[12pt]{article}

\usepackage{amssymb}

\newcommand{\ket}[1]{|#1\rangle}

\begin{document}     

 
\title{Self-Replicating Space-Cells and the Cosmological Constant} 

\author{Dirk~Vertigan\thanks{Department of Mathematics,  
Louisiana State University,  
Baton Rouge, Louisiana 70803-4918, USA.  
{\tt vertigan@math.lsu.edu}}}

\maketitle 

\begin{abstract} 

We consider what the implications would be if there were a discrete fundamental model of physics based on locally-finite self-interacting information, in which there is no presumption of the familiar space and laws of physics, but from which such space and laws can nevertheless be shown to be able to emerge stably from such a fundamental model. We argue that if there is such a model, then the familiar laws of physics, including Standard Model constants, etc., must be encodable by a finite quantity $C$, called the {\em complexity}, of self-interacting information $I$, called a {\em Space-Cell}. Copies of Space-Cell $I$ must be distributed throughout space, at a roughly constant and near-Planck density, and copies must be created or destroyed as space expands or contracts. We then argue that each Space-Cell is a self-replicator that can duplicate in times ranging from as fast as near-Planck-times to as slow as Cosmological-Constant-time which is $10^{61}$ Planck-times. From standard considerations of computation, we argue this slowest duplication rate just requires that $10^{61}$ is less than about $2^C$, the number of length-$C$ binary strings, hence requiring only the modest complexity $C$ at least $203$, and at most a few thousand. We claim this provides a reasonable explanation for a dimensionless constant being as large as $10^{61}$, and hence for the Cosmological Constant being a tiny positive $10^{-122}$. We also discuss a separate conjecture on entropy flow in Hole-Bang Transitions. We then present Cosmological Natural Selection II. 
\end{abstract}

\pagestyle{myheadings}
\thispagestyle{plain}
\markboth{DIRK~VERTIGAN}{Self-Replicating Space-Cells and the Cosmological Constant}


\section{Introduction}\label{S:intro}

This paper is directed primarily at theoreticians who take seriously the idea that there should in some form be a discrete fundamental model of physics based on locally-finite self-interacting information, in which there is no presumption of the familiar space and laws of physics, but from which such space and laws can nevertheless be shown to be able to emerge stably from such a fundamental model. We do not propose any specific such model, but instead, in Section~\ref{S:CFMP} we describe the presumed characteristics of such a model and in Section~\ref{S:CEFP} we describe the presumed characteristics of the emergence of familiar physics. From the assumption of such model and emergence we deduce in Section~\ref{S:SRSCCC} the phenomenon of Self-Replicating Space-Cells, and propose an explanation for the magnitude of the Cosmological Constant. In Section~\ref{S:MECHBT}, motivated by a question of Roger Penrose, \cite{RP}, we conjecture that in a Hole-Bang Transition, the Big Bang is generally formed with entropy minimized, and discuss some related topics. Finally in Section~\ref{S:CNStOC}, we discuss various conclusions, including implications for Cosmological Natural Selection, and the early, namely pre-Big Bang, history of the overall universe, and we propose Cosmological Natural Selection II. 

The arguments we present should be regarded as thought-experiments, or thought-theorizings, in an attempt to approach the scientific question ``What is actually true, and why?'' with the hope that such answers can ultimately be backed up by mathematically rigorous theory, and by observation and experiment. 

By {\em Planck units} we mean quantities of the form $G^i\hbar^j c^k$, such as {\em Planck time} $10^{-43}$ seconds and {\em Planck length} $10^{-35}$ meters, and we usually give quantities in these units, usually just as rough order-of-magnitude estimates. We use as Cosmological Constant $\Lambda=10^{-122}$, and write $N=\Lambda^{-1/2}=10^{61}$ for the corresponding time scale which we seek to explain. 

We sometimes write `space(time)' to mean `space and/or spacetime'. The word `string' refers to the kind in \cite{GJ}, rather than the kind discussed in \cite{LS3,PW}.

\section{Characteristics of a Fundamental Model of Physics}\label{S:CFMP}

We now describe the presumed characteristics of what we call a 
{\em Fundamental Model}, namely any form of a ``discrete model of physics based on locally-finite self-interacting information, in which there is no presumption of the familiar space and laws of physics, but from which such space and laws can nevertheless be shown to be able to emerge stably from the Fundamental Model''. The search for such models is alluded to in Lee Smolin's books, \cite{LS1,LS2,LS3}, which reference various original sources. We assume time is built into the model, with time proceeding locally and in discrete timesteps, although conceivably time could emerge from an even more fundamental model. Such a model could include a partial order of events, where $a\leq b$ means $b$ is in the causal future of $a$, similar to a discrete subset, with the $\leq$ relation, of a solution of General Relativity (GR), or the $1$-skeleton of a spin-foam from Loop Quantum Gravity (LQG). Spacelike slices could look like locally finite discrete objects such as labelled graphs, for example, a discretization of a spacelike slice of a GR solution, or a spacelike slice of a spin-foam, namely a spin-network from LQG. Locality of two events could be defined in terms of having a common event in the causal near future or near past. It is natural to suppose that the scale of immediate causality and immediate locality, are approximately Planck time and Planck length, as suggested by LQG which should be regarded as at least providing some guidance as to what a Fundamental Model could look like. In any case, directed and undirected graphs, or perhaps categories, can be used to represent causality and locality. If there were a list of candidates for Fundamental Models, then a strategy for choosing one could be to develop the notion of a {\em universal Fundamental Model}, namely one which could emulate all the others. 

We do not specify whether processes are deterministic, probabilistic, quantum or anything else. Anything mentioned could be replaced by linear, or maybe other, combinations of such things. Also we may treat any description of part of space or spacetime as a {\em patch} and then combine patches subject to compatibility conditions, thus including many-observer one-universe models, and holographic models, as alluded to in \cite{LS1,LS2,LS3}. A Fundamental Model should include a notion of equivalence, or perhaps also approximate equivalence, of ways that information represents physical reality, and such an equivalence could include, but not necessarily be limited to, equivalence arising from the action of some appropriate symmetry group. We try to make the argument in Section~\ref{S:SRSCCC} broad enough not to depend crucially on anything in this paragraph. The next paragraph lists some crucial assumed conditions. 

{\em Information} usually refers to classical or quantum information, but we allow it to refer to any mathematical description, provided the information can be quantified. Now, classical and quantum information can sometimes be quantified by any non-negative real number, of bits or qubits, but we shall round down to integer values, and call information of less than one bit, or similar unit, {\em trivial}, and otherwise {\em non-trivial}. We require the {\em Locally Finite Information (LFI) Condition}, namely that any finite region contains a finite quantity of information, which implies in particular, that it contains at most finitely many copies of any non-trivial piece of information. By the way, this condition, and the conclusions of Section~\ref{S:SRSCCC}, certainly do not hold for classical field theories, so discreteness plays a crucial role. We also require the {\em Necessary Interaction of Local Information (NILI) Condition}, namely that locally proximate information, must interact in the near future. Actually this condition may not need to be part of the Fundamental Model, but instead may emerge along with space. We also require the {\em Uniform Treatment of Information (UTI) Condition}, namely that the Fundamental Model has only one type of information and that the way it interacts is independent of whatever different kinds of things, such as space, laws, matter, etc., that the information may later emerge to represent.

\section{Characteristics of the Emergence of Familiar Physics}\label{S:CEFP}

Physics models often have a {\em trichotomy} of the information in their description, {\em Level-1} being a background space(time), {\em Level-2} being laws of physics, such as equations governing the interaction and behavior of fields or forces and particles, and {\em Level-3} being the actual fields or forces and particles interacting in space and time, subject to the laws of physics. Classical theories such as Newton's and Maxwell's clearly display such a trichotomy, and do so uniquely. In some theories there is some flexibility to shift between levels, for example a field that becomes, or is declared to be, constant, can be shifted from Level-3 to Level-2, and compact extra dimensions may fit in Level-2 or Level-1, or instead you could have more levels. By contrast, while General Relativity can be subjected to such a trichotomy, the levels are inextricably linked unlike the previous examples, and this is a main motivation for seeking a Fundamental Model that unifies the levels into a single form of information, satisfying the UTI Condition. 

The trichotomy idea remains useful for the purpose of discussing the emergence of familiar physics from the Fundamental Model, with the understanding that the trichotomy may be non-unique, and that the levels may be inextricably linked as with GR. 

An illustrative analogy is a universal Turing machine $T$ which on input string $AB$ simulates another Turing machine $T_A$ with input string $B$. We can regard $T$ as hardware, $A$ as software and $B$ as data. Or we can regard $T$ as hardware, and $AB$ as data. Or we can regard $T_A$ as hardware, and $B$ as data. 

It is also illustrative to consider classical and quantum circuits consisting of input and output (qu)bits and (quantum) logic gates, and also classical and quantum computers where some of the output loops back to some of the input, and some internal memory is added. There appears a dichotomy between information representing inputs/outputs/memory contents, and information representing logic gates, but it should all be treated as information interacting with itself. Such computers can serve as, perhaps approximate, models of a finite region of space, with information going both ways through the boundary surface, and information interacting inside. That part of the internal information that remains stable throughout time can be identified with the logic gates, while the varying part can be identified with the internal memory, a further dichotomy. On a related note a process may be apparently modelled by a mapping 
$(U,\ket{\psi})\mapsto(U,U\ket{\psi})$ and again there appears a dichotomy between the roles of $U$ and $\ket{\psi}$. Moreover the expression $U\ket{\psi}$ is not linear but rather is bilinear, so that the overall information interaction cannot be regarded as a linear process, though perhaps multilinear locally, and the linearity and multilinearity question may not even be meaningful as the Fundamental Model may not necessarily even have an operation of addition built in. So processes apparently modelled by $(U,\ket{\psi})\mapsto(U,U\ket{\psi})$ are another thing needing to emerge. And we reiterate that all the dichotomies in this paragraph are not built into the  Fundamental Model, but they can emerge from it. 

By the {\em O-universe} we mean all of our observable universe extended to everything, observable or not, coming from our Big Bang and all of its future, stopping at Black Holes, that is, our observable universe extended to everything in all of the spacetime approximated by a maximal solution of GR. The {\em O-space(time)} consists of the information representing space(time), with metric or equivalent information, of the O-universe, and we also call this information Level-1 when discussing emergence. The {\em O-Laws} consists of the information representing all the familiar laws of physics of the O-universe, including details of the Standard Model, particularly its constants, and the Cosmological Constant, and details of any future-discovered laws of this type, and we also call this information Level-2 when discussing emergence. Level-3 consists of the remaining information, which we call the {\em O-physical-activity}. 

An {\em A-universe} is defined the same way as the O-universe, but with respect to another observer, so the O-universe is just {\em our} A-universe. 
The {\em E-universe} will refer to everything that exists ever. 

We now state the assumptions that we assume in Section~\ref{S:SRSCCC}, from which we deduce the phenomenon of Self-Replicating Space-Cells and various corollaries. We assume that a Fundamental Model exists, satisfying the LFI, NILI, UTI Conditions from Section~\ref{S:CFMP}. We assume that this Fundamental Model models the O-universe, that is, it represents the whole history of the O-universe, in the form of some of its  self-interacting information, maybe via compatible patches. This requires that the abovementioned features such as all the details of Level-1 (O-space(time)), Level-2 (O-Laws), Level-3 (the O-physical-activity), can and do emerge stably from the Fundamental Model. 

It is worth noting all the various difficulties that can be encountered in trying to satisfy these assumptions. Firstly, it may be difficult to even find a Fundamental Model, of the type described in Section~\ref{S:CFMP}, and satisfying the LFI, NILI, UTI Conditions. It may be difficult to get anything like familiar space to emerge, after all, discrete mathematical objects almost never resemble a discretization of something continuous. Moreover, it has to persist through time, and model the geometry of GR, and to have the ability, when combined with other information, to behave like any region of the O-universe. The Level-2 information needs to represent the O-Laws, in a way that the O-Laws are constant, or as close to constant as is required by experimental and observational constraints, throughout space and time. The Level-3 information needs to interact with the Level-2 and Level-1 information so that it represents all the O-physical-activity, following the O-Laws in O-space(time). The Fundamental Model treats all information uniformly, but it nevertheless needs to process the information through time so that the three levels interact in exactly the right way, without otherwise mixing information between levels in a way that destroys the trichotomy. Actually, continuous models can also run into problems analogous to some of these, when they try to make their dynamics more unified, \cite{RP,LS3}. 

Some approaches to such difficulties are described or referred to in \cite{LS1,LS2,LS3}. We do not attempt to address these difficulties here. Instead we argue in Section~\ref{S:SRSCCC} what can be deduced whenever the abovestated assumptions are true. Logically, there is no problem for the argument from any situation where an assumption is false, nor from any situation where any additional assumption is imposed. Of course it would be desirable to ultimately demonstrate that these assumptions themselves are in fact true, as we believe they are. 

In our argument, whenever we refer to a feature of the real O-universe, or any physical model of it, we tacitly assume that we are also talking about the corresponding feature of Fundamental Model representation of the O-universe, and we are assuming there is such a correspondence. For example a volume $V$ will never be taken to be below the Planck scale.

\section{Self-Replicating Space-Cells and the Cosmological Constant}\label{S:SRSCCC}

We reiterate the assumptions we make, and note that some conclusions already appear in Section~\ref{S:CEFP}. We assume that a Fundamental Model exists, satisfying the LFI, NILI, UTI Conditions from Section~\ref{S:CFMP}. We assume that this Fundamental Model models the O-universe, it represents the whole history of the O-universe, in the form of some of its  self-interacting information.

Some of the conclusions we argue for are as follows. The O-Laws must be encodable by a finite quantity $C$, called the {\em complexity}, of self-interacting information $I$, called a {\em Space-Cell}. Copies of Space-Cell $I$, perhaps with some variation that does not affect the O-Laws, must be distributed throughout space, at a roughly constant and near-Planck density, and copies must be created or destroyed or merged, as space expands or contracts. Each Space-Cell $I$ is a self-replicator that can duplicate, again, perhaps with some variation, in times ranging from as fast as near-Planck-times to as slow as Cosmological-Constant-time which is $N=\Lambda^{-1/2}=10^{61}$ Planck-times. This requires that $10^{61}$ is less than about $2^C$, and the complexity $C$ is at least $203$, and at most a few thousand. 

It is illustrative to first consider some examples where neither the assumptions nor the conclusions hold, to see what aspects of the assumptions are crucial for the conclusions. 

An example violating discreteness and the LFI Condition: Suppose a model has an expanding manifold $M(t)$, and its O-Laws are modelled by a constant, or dynamic and nearly constant, compact manifold $L$ with fields on $M(t)\times L$. For any finite region of $M$ there is an uncountably infinite continuum of copies of $L$, and hence of the O-Laws, one copy for each point in the region, and as $M$ stretches, there is still an uncountably infinite continuum. There is no meaningful notion of replication in any such continuum based model. Discreteness is crucial. 

An example violating locality and the NILI, UTI Conditions: Suppose a model has just a single copy of the O-Laws, to apply to all of O-space(time), which is probably a common way to view models. Since there is always just a single copy, no replication occurs at all. This single copy scenario violates our assumptions, since if locality and the NILI, UTI Conditions hold, then the single copy of the O-Laws is just some information, which must be in immediate local proximity to all of space, from any spacelike slice, so all of space is locally proximate with itself and so must almost immediately interact with all of itself, effectively contracting space to a point. In fact a similar argument implies that there cannot be just a single copy of the O-Laws in a volume of space significantly above the Planck scale, assuming as usual that the Planck scale is the scale of immediate causality and locality. Thus the assumptions imply the crucial feature that Level-2 information representating the O-Laws is distributed roughly uniformly throughout O-space(time) at near-Planck density. 

An example violating the presence of gravitational phenomena: Suppose a universe has no gravitational phenomena, but is modelled by Quantum Field Theory (QFT) on flat spacetime. A Fundamental Model representation of such a universe may very well have features identifiable as Space-Cells distributed throughout space at a constant density, for all time. But there would be no replication. So gravitational phenomena, especially the expansion of space, are crucial. 

We will first present the argument, viewing the causality and locality structure as resembling a discretization of a GR solution that models the O-universe, with information localized in space and measured in bits. We later add comments to extend the breadth of the argument. 

We introduce the notion of a {\em $V$-Cell}, namely a region of space of volume $V$, of any shape within reason, and the Level-2 information it contains. Now the total Level-2 information is supposed to represent the O-Laws throughout all of O-space(time). So the Level-2 information throughout any $V$-Cell, should represent the O-Laws, and not some other Laws, nor should it represent something else entirely, nor should it be absent, and it should stably maintain these properties as time proceeds with all the Level-2, and 1 and 3, information interacting. As argued above, the Level-2 information representating the O-Laws is distributed roughly uniformly throughout O-space(time) at near-Planck density. It can be concluded that the amount of Level-2 information in a $V$-Cell, $C_V$ bits, is roughly proportional to $V$, say $C_V\approx\rho_2V$, though see variations of the argument below. Here we are just literally counting all bits, rather than what this repetitive information could be compressed to, and in any case, as discussed above, doing such a data compression amounts, in a sense, to contracting space to a point. 

We now consider a {\em $V(t)$-Cell} with volume $V(t)$ at time $t$. To start with, we will treat the O-universe as being approximately spatially homogeneous and isotropic with an increasing scale factor $a(t)$ as in \cite{SC}. A $V$-Cell at time $t_0$, which is comoving, see \cite{SC}, will grow as a function of time $t$ as $V(t)\approx Va(t)/a(t_0)\propto a(t)$, with $C_{V(t)}\approx\rho_2Va(t)/a(t_0)\propto a(t)$ bits of Level-2 information. 

Clearly, though see next paragraph, more Level-2 information is being produced. We argue that it is due to direct replication of Level-2 information occurring locally at the Planck scale. Firstly, it cannot be that new Level-2 information is being produced without any reference to pre-existing Level-2 information. If it were so produced, then effectively, the O-Laws would take zero bits to describe, which is highly implausible, unless one actually believes the O-Laws are somehow unique, namely $C=0$. In any case we later argue for a somewhat larger than zero value for $C$. Also, it cannot be that the new Level-2 information is being produced just by a general copying mechanism that copies everything, since this would lead to copying the wrong information, and would be too vulnerable to invasion by parasitic information, like a virus. So it must be that the pre-existing Level-2 information directly controls the copying of specifically itself to produce new Level-2 information. Then the question is, what groupings of Level-2 information serve as the basic units of replication. Since all processes occur locally at the Planck scale, the basic units of replication should be directly copyable at that scale. In addition, the basic units of replication should contain enough of the Level-2 information that all the rest of the Level-2 information is a copy of it, with some variation discussed later. So we define {\em Space-Cells} to be these basic units of replication, directly copyable at the Planck scale, and containing the required Level-2 information. We will add more about the definition, characteristics, interaction and replication of Space-Cells below. For larger $V$-Cells, having their information copied is explainable in terms of the replication of the Space-Cells they contain. Similarly copying of sub-parts of Space-Cells is explainable in term of replication of the Space-Cells they are contained in. Actually these sub-parts are also directly copied at the Planck scale, but don't contain all the Level-2 information, and they may depend on the rest of the Space-Cell for a copying mechanism. There is more discussion about replicators and replication in \cite{RD1,RD2,RD3,RD4,RD5,RD6,RD7}. 

We need to dismiss the possibility that instead of new Level-2 information being produced, there is a net flow of Level-2 information into the $V(t)$-Cell from neighboring regions. If the O-space is spatially finite, such as $S^3$, then this possibility clearly cannot occur for all such cells. In any case the Level-2 information is not only locally but also {\em globally} finite, and increasing in proportion to the scale factor $a(t)$, so again, new Level-2 information must be being produced. If the O-space is spatially infinite, such as ${\mathbb R}^3$, then we could argue that the situation is locally very similar to the spatially finite case, justifying the same conclusions. On the other hand, the spatially infinite case, though {\em locally} finite, would globally have a {\em countably infinite} number of Space-Cells. It seems such a situation could not arise within the overall theme of this paper, the question being, how do you get to a countably infinite number of Space-Cells in the first place. So there is no particular need to argue for the spatially infinite case anyway, as it appears we need finite space. Compare these cases to continuum models which are both locally and globally {\em{\bf{\em un}}countably infinite}. 

We have now argued for the phenomenon of Self-Replicating Space-Cells, but we first continue discussing them before moving on to the Cosmological Constant. 

Let {\em complexity} $C$ be the number of bits of information $I$ in a Space-Cell. There can be some variation in this information, without affecting the fact that it represents the O-Laws. We take into account this possibility, in a simplified way, by allowing that a certain $D$ of the $C$ bits can differ between two Space-Cells, and these $D$ bits can also vary in time, while the remaining $C-D$ bits are fixed. Apart from allowing this variation, these $D$ bits can play some other roles as we now list. Recall from Section~\ref{S:CEFP}, we can regard the self-interacting information $I$ as encoding a computer, with the $C-D$ bits encoding the logic gates, while the $D$ bits are the internal memory. Actually, this computer could also operate on the Level-3 information. The contents of the $D$ bits, in many Space-Cells over a region in space, could form patterns that play a role in the emulation of the O-Laws, although the potential for this to happen should be implicit in the information in one Space-Cell. The $D$ bits could also play a role in some type of error correction, or in some way making Space-Cells more stable and robust. 

It might seem that roughly speaking, a $V$-Cell can be partitioned into about $n(V)\approx C_V/C\approx\rho_2V/C \approx V/V_0\propto V$ Space-Cells, where $V_0=C/\rho_2$. It is reasonable to suppose that $V_0$ would be somewhere between $1$ and $C$ Planck volumes. The quantity $n(V)$ is reasonable as a measure of information. However a $V$-Cell cannot generally be partitioned neatly into Space-Cells. One complication is that two Space-Cells can partially overlap, and in fact this is inevitable, since such a scenario could represent an intermediate stage of one Space-Cell splitting into two. Also, parts of two or more Space-Cells may constitute another Space-Cell. These complications don't affect our arguments above, and in any case a more technical approach to counting Space-Cells could later be developed as needed. They are mentioned to convey how it is that Space-Cells cover all the Level-2 information. 

The above use of $V$-Cells and the scale factor $a(t)$ were for illustrative purposes. It should follow more generally, for arbitrary GR spacetimes, that Space-Cells maintain a roughly constant and near-Planck density, and so Space-Cells must be replicated or destroyed or merged, to maintain this. Ultimately it should be shown that all aspects of the metric, or equivalent information, can be emulated by the behavior of Space-Cells within O-space, namely the Level-1 information, but we don't need to further delve into that question here. 

We have argued that Space-Cells are replicators, but they are probably not what we will call {\em classical replicators} by which we mean an entity that maintains its identity between replication events. By contrast, in the Fundamental Model, bits and larger clusters of information, such as Space-Cells and parts thereof, follow what are best described as lightspeed Planck-scale zig-zag paths, and such a path can be chosen to approximately follow essentially any worldline. Furthermore interactions will commonly take the form of a {\em pair} of bits or clusters of information or Space-Cells $w,x$ which merge, interact and then yield another such pair $y,z$. But each one of $w,x$ is not necessarily identified or associated with either one of $y$ or $z$ any more than the other, and a worldline through $w$, say, could equally well then pass through $y$ or through $z$. So there can be many more $2$-Space-Cell$\longrightarrow2$-Space-Cell interactions than there are $1$-Space-Cell$\longrightarrow2$-Space-Cell replications, so that Space-Cells don't seem to be classical replicators, though they are definitely replicators. 

We consider now the possibility that the causality and locality provided by the Fundamental Model representation of the O-universe, may not simply look like a discretization of a GR solution that models the O-universe. Instead they could be more abstractly related so that the Fundamental Model representation may even have interactions that are not manifestly local in the O-universe.  Of course, it must be consistent with the above assumptions, and hence with theory and experiment. Also, as mentioned in Section~\ref{S:CFMP}, there may be variations, in the Fundamental Model or in the emergence, involving compatible patches, including many-observer one-universe models, and holographic models. In all cases, Space-Cells are clusters of information in the Fundamental Model representation itself. Moreover, the abovementioned volume, scale factor $a(t)$ and in fact the whole metric, and also the localization and quantification of information of any type, and the associated counting of Space-Cells, all need to be expressed directly in the Fundamental Model representation. Nevertheless, if these models appropriately emulate GR, and the conditions regarding information hold, then it is reasonable to expect that the arguments extend to these cases. 

Space-Cells need to be replicated or destroyed or merged, as O-space expands or contracts, locally or globally. Space-Cells being destroyed, seems a straightforward process, and if time-symmetry is insisted upon, so that instead Space-Cells should merge 2-to-1, then we can just consider the time-reversal of replication. For expansion and the corresponding replication, we want to find the fastest and slowest rates of expansion, and consider if repeated replication of Space-Cells can plausibly occur at the required rate. For this purpose it is sufficient again to consider the scale factor $a(t)$ and the associated {\em Hubble parameter} $H(t)=\dot{a}(t)/a(t)$, \cite{SC}. The associated time over which one replication is to occur, is called the {\em doubling time} $r(t)=1/H(t)=a(t)/\dot{a}(t)$, ignoring small factors such as $\ln 2$ and $3$ etc. In the O-universe, the largest $H(t)$ and smallest $r(t)$ are early after the Big Bang, especially if there is {\em inflation}, \cite{SC}, which we can accommodate, but may or may not require. Since replication is potentially a simple process, a $C$-bit Space-Cell, to replicate as fast as possible, may just require between $1$ and not much more than $C$ Planck timesteps to replicate, and it can certainly replicate at any slower rate, as needed, up to a maximum slowness, see below. This should be able to accommodate the required doubling time for any proposed inflationary scenario, and can certainly accommodate the other early epochs after the Big Bang. 

At the other extreme, the smallest $H(t)$ and largest $r(t)$ are for sufficiently large $t$, when the O-universe is approximately a de Sitter universe, whose expansion is determined purely by the positive cosmological constant $\Lambda=10^{-122}$, in what is traditionally viewed as `matter-free empty space'. In this case, $H(t)\approx1/N=\Lambda^{1/2}=10^{-61}$ and $r(t)\approx N=\Lambda^{-1/2}=10^{61}$. We seek to explain how replication can be held to such a slow rate. The explanation is essentially purely in terms of Level-2 information. Note that, the Level-1 information determines which Level-2 information interacts and when, but the upshot of that is just that the Level-2 information interacts locally in O-space(time) as expected. The Level-3 information is absent in this `matter-free empty space'-scenario, so it contributes nothing, except it could possibly be allowed to contribute some generic randomness, which could have been present in the model already anyway. 

Recall that, amongst other things, we could regard a Space-Cell as encoding a computer, with the $C-D$ bits encoding the logic gates, and with the $D$ bits being internal memory. We will use some basic concepts from computing theory, \cite{GJ}. A computer with $D$ memory bits has $2^D$ states, disregarding other variable states within the computer that may contribute a small factor. If it is deterministic, and it runs for more than $2^D$ timesteps, without external input during those steps, then it must have repeated a state and so would be in a recurring loop. For such a computer to complete a task and stop, it must do so in at most $2^D$ timesteps. If a task can be done quickly, and one wants to make it take longer, then one can delay the task by simply counting out some extra timesteps, using the memory to store the count, but the overall upper bound of $2^D$ timesteps nevertheless applies. 

By analogy, and just as an illustrative oversimplification to start with, one way a Space-Cell can replicate about as slowly as possible, given that it can replicate very fast, is to simply count through about $2^D$ Planck timesteps, and then replicate. But similarly to above, it cannot take longer than this. This suggests the inequality $N=10^{61}\leq 2^D$ so that $\log_2(N)\approx202.6\leq D$, although $D$ may be somewhat larger. So far this treats a Space-Cell as being isolated, and as just computing internally on its $D$ memory bits. It is more realistic to view it as a Space-Cell worldline, see earlier this section, always interacting with external information. In the absence of a more sophisticated model, we will treat the effect of this external influence as putting the Space-Cell, at each Planck timestep, randomly and equiprobably into one of its $2^D$ internal states. If there is some non-empty subset of these states that trigger a replication event, then at any timestep the probabilty of a replication starting is at least $2^{-D}$. Averaging over a population of Space-Cells again yields the $N\leq 2^D$ inequality. Now it may well be that one could contrive a sequence of inputs to the Space-Cell to yield an arbitrarily long delay before a replication, but such a scenario does not realistically model a Space-Cell in an environment of nothing but interacting Space-Cells. The simplified idea of a state `triggering' a replication event could be further sophisticated by knowing all the intermediate stages of one Space-Cell splitting into two Space-Cells, and knowing all their possible internal states, and knowing how external information input causes changes between these stages and states. In any case we believe the stated inequality is robust to further elaborations of the model. So, 

\large
\begin{equation}
N=\Lambda^{-1/2}=10^{61}\leq 2^D<2^C, {\rm\ so\ that\ }202<D<C. 
\end{equation}
\normalsize

One thing this explains is not so much the specific value $N=10^{61}$, but rather it explains how a number as large as $N=10^{61}$, and hence the Cosmological Constant $\Lambda=N^{-2}=10^{-122}$, can arise from a process involving a logarithmically smaller quantity of interacting information. The experimentally measured tiny positive Cosmological Constant $\Lambda=10^{-122}$ in turn provides evidence for a fundamentally discrete physics model, and provides evidence for Self-Replicating Space-Cells with a relatively small complexity $C$ of at least a few hundred, and maybe thousands but not much more, see below. This should be compared to other proposed explanations of the Cosmological Constant. 

Now the complexity $C$ is at least about $200$ bits, though it could be somewhat more. We suggest a common sense estimate for an upper bound for $C$. If for example, one wanted to specify about $20$ dimensionless constants to $30$ signifiant decimal places, namely about $100$ signifiant binary places, one would need about $2000$ bits. It would be reasonable to add one bit for each binary order of magnitude each constant differs from unity. Some information would be needed to encode the other details of the O-Laws. So a few thousand bits certainly seems enough. Actually, the encoding of the O-Laws in a Space-Cell could be quite different from an explicit literal encoding of constants in binary. Instead the values of the constants quantify aspects of processes encoded by the Space-Cell, and these constants could be related in quite a complicated way to the information in the Space-Cell. Nevertheless, it still seems a few thousand bits should be enough. 

Now we haven't actually given an argument for the size of $C-D$, so although it seems unlikely, it could conceivably be much smaller than $D$ and $C$. A rough analogy would be that an extremely short program could be made that runs for an extremely large number of steps, for example $n=3!!!!!!$ steps, but it would still require at least $\log_2 n=\log_2 3!!!!!!>>3!!!!!$ memory bits to run. So there is an interesting question as to the relative sizes of  $C-D$ and $D$. We actually expect they do not dramatically differ, and that the numbers $C-D$, $D$, $C$ are all in the hundreds or thousands. In any case, a Space-Cell consists of $C$ bits of information, so that is the appropriate parameter to call the complexity.

\section{The Minimal Entropy Conjecture for Hole-Bang Transitions}\label{S:MECHBT}

This section is more conjectural, and is somewhat independent of the previous sections, but all the sections will lead into the next section. 

In \cite{RP}, Roger Penrose explains that our Big Bang started in a state of very low entropy, and he poses the question of why this should be so, the idea being that there should be some explanation, through some yet-to-be discovered physics, rather than it just being sheer luck. He explains that the main contribution to entropy is gravitational for which a uniform distribution of matter corresponds to low entropy, and a non-uniform distribution of matter corresponds to high entropy, while other contributions to entropy are much smaller, and have this correspondence reversed. 

Various theorists propose the idea that from a Black Hole, also through some yet-to-be discovered physics, one or more Big Bangs may be produced, each starting a new A-universe. We call these {\em Hole-Bang Transitions} or {\em HBT's}, and we join the conjecture that they are a real phenomenon. They are crucial in Cosmological Natural Selection, \cite{LS1}, which we discuss in Section~\ref{S:CNStOC}. 

It is natural for us to conjecture that in a Hole-Bang Transition, the Big Bang is formed with entropy minimized, perhaps subject to various parameters of the Big Bang, yet again, through some yet-to-be discovered physics. If this conjecture could be proved it would provide an answer to Penrose's question. And it seems that Cosmological Natural Selection would need this conjecture to be true. 

For a high entropy Black Hole to produce a low entropy Big Bang, without violating the 2nd Law of Thermodynamics, there would need to be entropy removed on the Black Hole side. By analogy, consider a low entropy crystal being formed with the release of high entropy heat. Now entropy is removed from a Black Hole by Hawking radiation, although that process is very slow. However, inside the Black Hole, perhaps further in than the region modelled by GR, there may be a much more rapid process that removes entropy leaving a low entropy core. From this low entropy core a new low entropy Big Bang could be formed, perhaps requiring some additional process, such as an inflationary expansion. 

An important question is in what way does the new A-universe formed depend on what happens to the Black Hole from which it formed, and vice versa. 

It is interesting to consider what a minimal entropy Big Bang should look like. For minimizing entropy, the gravitational contribution to entropy favors a uniform distribution, while the other much smaller contributions to entropy favor a non-uniform distribution. These contributions may be of a form such that the total entropy is minimized by an almost but not perfectly uniform distribution, consistent with, and maybe explaining, what is observed. 

When considering discrete models, an even much bigger contribution to entropy, than the gravitational contribution, will pertain to how much a discrete structure resembles a discretization of space. Generally, an arbitrary discrete structure will have a much higher entropy than a discrete structure which resembles a discretization of space. This adds an even bigger component to Penrose's observation that our Big Bang started in a state of very low entropy. Moreover, it unifies his question about the low entropy Big Bang, with the important question of why there is anything resembling space at all, rather than something else utterly different. The Minimal Entropy Conjecture for Hole-Bang Transitions, if proved, could answer both questions in a unified way. 

Actually, explaining the emergence of space in this way would require that a Big Bang with the usual space is yielded from something analogous to a Black Hole in something not resembling space. A plausible notion for capturing the right concept would be to define, in the context of self-interacting information from the Fundamental Model, the notion of a {\em causal bottleneck}, in other words, to generalize the notion of a Hole-Bang Transition to a form that does not involve space in the definition.

One interesting possibility arises from contemplating the ever increasing entropy of an A-universe. A variation of the Heat Death scenario could be {\em The Big Unravelment} in which, over time, the discrete structure of A-space unravels into a discrete structure bearing no resemblance to a discretization of space, and whose `diameter' is the log of its `volume'.

\section{Cosmological Natural Selection II and Other Conclusions}\label{S:CNStOC}

It is important to note that the argument for the phenomenon of Self-Replicating Space-Cells just used ideas from physics, mathematics and computer science, starting with the assumption that our O-universe is modelled by a Fundamental Model. This argument did not assume in advance that any kind of replicators were involved, nor did it assume any other analogies with biology. Nevertheless once we have these conclusions, the fact that there are analogies with biological themes becomes completely obvious. It is appropriate to explore just how deeply the analogies run, as part of the quest to answer the scientific question ``What is actually true, and why?''. 

As discussed in \cite{RD1,RD2,RD3,RD4,RD5,RD6,RD7}, whenever there are replicators, and some other virtually automatic conditions, there is an evolutionary process. We noted in Section~\ref{S:SRSCCC}, that Space-Cells don't seem to be classical replicators, though they are definitely replicators, and there is nevertheless certainly an evolutionary process. In fact Earth's earliest pre-biological replicators, in the perhaps tens of millions of years of evolution culminating in the first simple biological cells, were also probably not classical replicators, due to a prevalent horizontal transmission of information at those earliest stages. 

Now Section~\ref{S:SRSCCC} also did not make any particular assumptions about what there was beyond the O-universe. But the question arises as to what kind of history there was, all of it modelled by a Fundamental Model, leading to the O-universe but starting from a state totally lacking in all the features that had later emerged by the time the O-universe was formed. It is natural to conclude that Self-Replicating Space-Cells have an evolutionary history through a line of replicator ancestors some of which may have existed in very different environments to the O-universe. 

Also there needs to be a Big Bang producing mechanism, and Hole-Bang Transitions are the most compatible with what we describe, as will soon become clear, and we won't consider other such mechanisms. 

Another question arises as to why we have our specific type of O-Laws, and corresponding type of Space-Cell, rather than some other Laws. An obvious part of the explanation is that Space-Cells are the products of an evolutionary process, which then leads to the question of what is it about our type of Space-Cell that makes it the kind of thing that such evolutionary process would yield. Many details to such an explanation are already to be found in Lee Smolin's proposal of {\em Cosmological Natural Selection (CNS)}, \cite{LS1}, in which, in our paper's terminology, the replicators are whole A-universes with their A-Laws, and replication occurs via Hole-Bang Transitions, yielding A-universes with slightly modified A-Laws. The overall process produces a population of A-universes dominated by those that produce many Black Holes, and hence many offspring A-universes. While CNS did not involve anything like Self-Replicating Space-Cells, once Space-Cells are brought into the picture, the explanation applies to them, and also Space-Cells become part of the explanation. It could also be argued that conversely, from the phenomena of Self-Replicating Space-Cells and Hole-Bang Transitions, one could have deduced the basic idea of Cosmological Natural Selection as a corollary. Space-Cells provide a mechanism for encoding the A-Laws and for their passage to the offspring A-universes, the latter obviously requiring that Space-Cells can in fact pass through a Hole-Bang Transition. Now it could be that during HBT's, Space-Cells may be modified, leading to modification of the A-Laws, just as in CNS. Additionally, Space-Cells, as evolutionary objects, already have their own variation, and even within an A-universe, there can be some slight variation, subject to observational constraints, and whatever future theory may explain that. This provides an additional source of variation to the A-Laws that does not require HBT's. 

Alternatives to CNS have various weaknesses as discussed in \cite{RD7,RP,LS1}, and we won't consider such alternatives. 

It seems obvious that Self-Replicating Space-Cells and Cosmological Natural Selection are two aspects of the same evolutionary process, and we call the combination, and the whole evolutionary history of Self-Replicating Space-Cells through a line of replicator ancestors, {\em Cosmological Natural Selection II (CNS2)}. Note that neither of these proposals assumed the other. But any theory, experiment, or observation that supports one, would generally tend to support the other. 

Many details are given in \cite{LS1} about what kind of evdience can support CNS, although of course direct experiment and observation are limited to just one of its replicators, namely the O-universe itself. By contrast, Self-Replicating Space-Cells fill the O-universe at near-Planck density, so there are plenty of such replicators to observe and experiment with, at least in principle, and sufficiently compelling evidence could justify extrapolation beyond the O-universe. Probably many theoretical and other developments would need to precede any proposal for observation and experiment that could provide evidence favoring Self-Replicating Space-Cells over alternative physical models. But already, as described in \cite{LS2}, experiments are underway seeking evidence for Planck scale discreteness of space, which would at least lend weight to the concept of Self-Replicating Space-Cells. 

One striking analogy is that, just as multi-celled organisms are built from biological cells, the much huger A-universes are built from the much tinier Space-Cells. To borrow a phrase from \cite{RD1}, A-universes can be viewed as {\em survival machines} for the Space-Cells. We now consider to what extent other such analogies may be drawn. Depending on the topic, a Space-Cell can be regarded as being analogous to anything ranging from the genome up to the whole biological cell. In all cases it should be emphasized that a Space-Cell is not just passive information, but should always be regarded as self-interacting information, interacting internally and with its surroundings. Also a Space-Cell is vastly simpler than a biological cell, and presumably so is its replication mechanism. Space-Cells can overlap in many ways, and not just when part way through a replication. Biological cells clearly partition space with distinct boundaries. In Section~\ref{S:SRSCCC} we discussed how there can be many more $2$-Space-Cell$\longrightarrow2$-Space-Cell interactions than there are $1$-Space-Cell$\longrightarrow2$-Space-Cell replications, so that Space-Cells don't seem to be classical replicators, though they are definitely replicators. Biological cells, genomes, genes, and other biological replicators, seem more like classical replicators, as they generally maintain their identity between replication events, albeit with some exceptions. 

It seems that A-universes, unlike biological organisms, don't interact with anything else, except via Hole-Bang Transitions, although the possibility of other forms of interaction should not be totally dismissed. However Space-Cells definitely interact with each other, and there can certainly be a selection process between different types of Space-Cells, closely analogous to natural selection in biology. Now the O-universe, and any A-universe, with almost exactly uniform A-Laws throughout, seems to have just one type of Space-Cell. Firstly, this is, in any case, closely analogous with copies of the genome within a single biological organism. Secondly, in principle there will nevertheless still be some variation on which a selection process can work. 

An interesting scenario is that of a {\em var-A-universe} which is like an A-universe, except that it has a mixture of different Space-Cell types, and also differences between such types and mixtures in different regions. The var-A-Laws would similarly vary.  There are many possibilities for the consequences of interactions between different Space-Cell types. They could destroy each other, or even the structure of space itself, or they could interact and form new Space-Cell types, or they could  form various kinds of combinations, or they could barely interact at all and just disperse osmotically, or do something else utterly different. There could be local patches of uncontrolled inflation, {\em space cancer}, and there could be parasitic information using a Space-Cell's copying mechanism, {\em space viruses}. In any case, if the situation remains fairly stable, then some of the following processes may occur. Only Space-Cells surrounding a Black Hole will fall into it, so the new var-A-universe formed will only have a local sample of Space-Cells, albeit still possibly a mixture, and regional differences will be filtered out. With  a local mixture of Space-Cell types, a selection process may drive some types to extinction. Finally different Space-Cell types may enter into a symbiotic relationship, and may permanently hybridize into what becomes identifiable as a single Space-Cell type. By such processes, a var-A-universe may, after a number of generations, have A-universes as its descendants. Our O-universe, and A-universes in general, may have had var-A-universes as their ancestors. And a corresponding line of descent follows for the Space-Cells they contain. Of course it would be interesting to go further back to the initial emergence of space, and the appearance of the first simple replicators. 

The question now arises as to whether the Space-Cells in our O-universe, and in A-universes in general, could be symbiotic hybrids of simpler Space-Cells. They could be. And this possibility is included in the definition of Space-Cell with its $D$ out of $C$ variable bits. Such a hybrid can be described in terms of its sub-parts and their relationships in the required way. The variation captured in the $D$ bits can include the variation within each sub-part, the variation in how they are related, and possibly even some variation in the actual list of sub-parts. In any case, it seems appropriate to regard certain sub-parts of Space-Cells as replicators, analogous to genes in a genome, recalling though that Space-Cells are much simpler with less room for sub-parts. 

It is explained in \cite{RD1} how multi-celled organisms will generally evolve so that their offspring start as a single cell, rather than as a larger cluster of cells detatching from the parent. By contrast, in CNS2, the new A-universe formed in a HBT inherits many Space-Cells from its parent. One possible explanation for the difference is that, as a technicality, the argument in \cite{RD1} seems to make use of the fact that biological cells are classical replicators, which does not apply to Space-Cells. Another difference is that offspring A-universes need to be provided not only with Space-Cells, but with an entire environment, namely A-space. Finally, there are expected to be universal constraints on what is physically possible, beyond the parameters that Space-Cells can vary, so some conceivable offspring-universe-producing alternatives to HBT's might not actually be possible. 

As just alluded to, some aspects of our familiar physics could be varied by having different Space-Cells, while other aspects may be universal. On the one hand, it may be possible to vary the Standard Model constants, and some of its other details, and the Cosmological Constant, by having different Space-Cells. On the other hand, it may be that with some built-in symmetry principle in a Fundamental Model, once the right type of space can be shown to emerge, then it could be shown that familiar symmetry principles such as general covariance and gauge invariance would automatically follow, yielding various other consequences as well. 

It is envisioned that some generalization of the 2nd Law of Thermodynamics would also be universally applicable. This allows, of course, local regions of low entropy. Moreover, this allows for mechanisms that produce such regions of low entropy, at the expense of their surroundings as usual, such as Hole-Bang Transitions. This makes it possible to permanently evade a global Heat Death scenario for the entire E-universe. Thus, while a Heat Death scenario, including possibly The Big Unravelment, see Section~\ref{S:MECHBT}, is expected for an A-universe, rejuvenation is provided by Hole-Bang Transitions, creating new low entropy A-universes. 

It could reasonably be asked why Space-Cells shouldn't simply replicate as rapidly as possible in a permanent inflationary scenario. We conjecture that this may simply not be possible, as such a scenario may rapidly lead to a Heat Death scenario, including possibly a Big Unravelment, see Section~\ref{S:MECHBT}, in which the structure of space itself unravels, placing a limit on the sustainability of any inflationary scenario. Nevertheless a controlled inflationary epoch can be accommodated in CNS2 if needed, as long as it doesn't last too long. Whether there is an inflationary epoch, and what its parameters are, is something that is to be evolutionarily optimized in CNS2.  

In CNS, adjustable parameters take values near a local optimum for maximizing production of offspring A-universes, and in CNS2 the same holds for Space-Cells. There are many details in \cite{LS1} of observational and theoretical support for this, which are also applicable to CNS2. Just as in biology, the situation can be very complicated, and the explanation or prediction of optima cannot always be expected to be transparently obvious. In any case, the contention is that our O-Laws including a Cosmological Constant $\Lambda=10^{-122}$, and a Space-Cell doubling time of $N=\Lambda^{-1/2}=10^{61}$ Planck times, is near-optimal for our O-universe, and is superior to an attempted permanent inflationary scenario, with its near-Planck doubling rate. 

Recall the discussion in Section~\ref{S:SRSCCC}, explaining the reasonableness of a Space-Cell doubling time as slow as $N=10^{61}$. Essentially time periods of $N=10^{61}$ steps can simply arise from dynamical processes involving as few as $\log_2(N)\approx203$ bits. We found that we need $N=\Lambda^{-1/2}=10^{61}\leq 2^D<2^C,$  so that $202<D<C$, and suggested that maybe $C$ and $D$ would be a few thousand at the most, making Space-Cells relatively simple evolutionary objects. The information determining the various constants, namely those in the Standard Model, and the Cosmological Constant, and perhaps some other constants, such as the parameters of a controlled inflationary scenario, if any, would be implicitly encoded in the structure of the Space-Cell. It is reasonable to expect that the details of this encoding, and of the replication, variation and selection processes, allow for the optima to be approached as described in CNS. Thus, not only can time periods of $N=10^{61}$ steps easily arise from relatively small Space-Cells, but also variation to nearby values should be easily obtainable. We empasize that there is no need to explain $\Lambda=10^{-122}$ in terms that require fine tuning to $122$ decimal place precision.

\end{document}